\documentclass[aps,amssymb,twocolumn]{revtex4}
\begin{document}
\title{
Common origin of no-cloning and no-deleting principles - Conservation of information
}

\author{
Micha{\l} Horodecki$^{1}$, 
Ryszard Horodecki\(^{1}\), Aditi Sen(De)$^{1,2}$, and Ujjwal Sen$^{1,2}$}

\affiliation{\(^1\)Institute of Theoretical Physics and Astrophysics, University of Gda\'nsk, 
         80-952 Gda\'nsk, Poland\\
\(^2\)Institute for Theoretical Physics, University of  Hannover, D-30167 Hannover, Germany}

\begin{abstract}
\
We discuss the role of the notion of information in the description of physical reality.  
%We show that \emph{independent of any theory}, 
%the  no-cloning and no-deleting principles follow from the law of 
%conservation of information, and whether two copies contain a different amount of information 
%than a single copy. 
We consider theories for which dynamics is linear with respect 
to stochastic mixing.
We  point out that the no-cloning and no-deleting principles emerge in any such theory,
if law of conservation of information is valid, and two copies 
contain more information than one copy. We then describe the quantum case from 
this point of view.
\vskip1cm
This paper  is dedicated to Asher Peres on the occasion of his seventieth birthday. 
\end{abstract}

\maketitle

\def\com#1{{\tt [\hskip.5cm #1 \hskip.5cm ]}}

The fact that Nature allows us to describe itself mathematically, seems even more 
astonishing than the laws of Nature themselves. In particular, quantum formalism is enigma,
which in spirit of G\"odel's theorem can not be explained by itself. The lack
of clear relation between the description and reality, brings about difficulties in 
the interpretation of quantum formalism. As one knows, any attempt of  
objectivisation of the latter leads to a
number of paradoxes \cite{1Peres}. This gap between formalism and reality has, in particular, its reflection
in Asher Peres's phrase: ``the physics is what  physicists do in laboratories" \cite{1Peres}
and ``entanglement is a trick the quantum magicians use to produce phenomena, that 
can not be imitated by clasical magicians" \cite{2Bruss}. But again, why quantum magicians are better 
than their classical counterparts?   This question
forces us to adopt the primitive notions which are autonomic  with respect to the formalism.
In this context two notions seem to be relevant: information and 
informational isomorphism \cite{3RPM-IBM}.

The recent discoveries \cite{4BB84,5PeresW,6Ekert,7BWiesner,8telepor} concerning processing of 
information in the quantum regime \cite{23foot}
convince us more 
and more that Landauer's slogan, ``Information is physical!" \cite{11Landauer}, is not empty \cite {Weinfurter}. 
The idea that the notion of information should be regarded as a fundamental ingredient in a
physical  theory was proposed in different contexts \cite{12RH,13MRH98,14Zeilinger,16MPRcompl,17Jozsa}. 
However it is not quite clear, 
what the term ``physical'' means, in the above context. 
%means in the above context the term "physical". 
In fact, there are two opposing 
pictures of  information: i) subjective,  according to which  information  
represents  {\it knowledge}, ii) objective, which treats information (just like energy) as a {\it property} of 
the physical system.  This cognitive duality can be surmounted by postulating  
that any consistent description of Nature, is a sort of isomorphism between the laws of 
Nature and  their mathematical representation.  According to this view \cite{3RPM-IBM}
(called informational isomorphism), although  no notion itself is  
reality, yet it reflects physical reality. Then any theoretical structure, although 
is not a real thing, is an isomorphic image of the existing reality.
In this sense, information  can be treated as physical, and it is natural 
to ask: What are the fundamental consequences  of such statement? 
The problem is by no means  trivial, as it concerns the properties of the 
quantity that we regard as physical,  which nevertheless, manifests itself in a way  
that is far from intuition. 

One of the basic properties that a physical quantity can exhibit is conservation  of the quantity 
under some physical process.  Historically,  the  principles of conservation 
played a central  role in the development of the modern physics on both the  
classical and quantum levels of the description of physical reality. For example, 
Pauli discovered neutrino in the beta-decay process, basing only on the conservation 
of energy-momentum. By analogy, it is reasonable to postulate  the conservation of 
information as a principle,  and raise  it up to a paramount law of nature.  In this paper 
we adopt this view and investigate its consequences,  in  the context of processing  of 
classical and quantum information. 
%Basing on the principle of conservation  of 
%information and the von Neumann  entropy as a measure of information we derive 
%the both no-cloning  and no-deleting theorems.
We will consider any theory for which the dynamics is linear with respect to stochastic mixing.
We  point out that the no-cloning and no-deleting principles emerge in any such theory, 
if the law of conservation of information is valid, and two copies 
contain more information than one copy. We then describe the quantum case from 
this point of view, using von Neumann entropy as a measure of information.

We should emphasise here, that we do not want to derive 
the no-cloning and no-deleting theorems from unitarity. This has already been done, 
and is not the purpose of our paper. Rather, we would like to 
present a proof, that will separate the argument into two parts:
i) the part independent of quantum mechanics, 
referring solely to conservation of a physical quantity (which is information), 
and ii) the quantum mechnical part,  where the quantum mechanical 
information is evaluated.

To explain the relevance of such an approach, let us recall  the principle 
of conservation of energy. The principle of 
conservation of energy was discovered in the XIX century, in the context 
of trials to build a {\it perpetuum mobile}. The principle was thus born 
on the ground of classical physics. But it remains a fundamental principle in quantum 
physics. The  notion of energy and  its conservation survived 
the revolutions in physics - both from classical to quantum, and from Galilean 
to relativistic. Thus this notion seems to be more fundamental 
than specific theories, and it is treated as a basic property of 
physical systems. It is therefore good to understand some phenomena 
solely on the ground of a principle, such as energy conservation, rather
than to derive them within some theory.

The aim of this paper 
is to tilt our understanding of no-cloning and no-deleting theorems
in a similar direction, but with respect to {\it information}. We will simply state that one cannot 
clone or delete, in any theory if two copies have more information than one copy in that theory.
Then we check that it is the case in quantum mechanics, unlike in classical 
world, where two copies represent exactly the same amount of information.

\def\com#1{{\tt [\hskip.5cm #1 \hskip.5cm ]}}

\newcommand{\tr}{{\rm tr}}

Let us first specifically state what we assume about the operations that can be performed over a physical system:
\begin{itemize}
%\item[(i)] The dynamics of the system is linear on the level of pure states (this can of course be easily generalised 
%to mixed states);
%\item[(i)] The system is always in a pure state (i.e. complete information about the system is 
%available);
\item[(i)] Enlargement of the system is allowed. I.e. 
addition of another system (containing no information about the 
original system) is allowed;
\item[(ii)] The dynamics of the closed system is linear with respect to  stochastic mixing. I.e. the dynamics is 
linear over  the stochastic mixture \(\varrho = \sum_i p_i \varrho_i\) of an ensemble \(\{p_i, \varrho_i\}\).
\end{itemize}   
%Now, we adopt the conservation of quantum information as a basic principle, which can be 
%as a 
(An ensemble \(\{p_i, \varrho_i\}\) denotes a source producing 
a state \(\varrho_i\) with probability \(p_i\).)

To obtain no-cloning and no-deleting from the law of conservation of information, we ask the 
following question: In a given physical system, is there a difference 
 between the information content in a single copy with respect to that in two copies?

In a cloning process, we produce two copies of the input. Additionally there can be some ``garbage'' produced 
in the output. We may discard this additional part into the environment.  If in a cloning 
process, this output (two copies plus possible garbage) has more information than the input (single copy), then
we have a violation of  
the law of conservation of information. 

In a deleting process, we do not allow additional garbage to be produced in the environment. 
So if two copies have more information than a single copy, we have no-deleting 
implied by the law of conservation of information.

Note that it is natural that we do not allow discarding part of the system (the ``garbage''),
 as a valid operation in considerations
of  no-deleting. When we are trying to show that reducing
 information from the system is not possible, we 
cannot afford to be ``careless'' and allow throwing out parts of the system. On the 
other hand, in considerations with no-cloning, when we are trying to show that 
production of information in a system is not possible, we must allow discarding. 
After all,
throwing out part of a system cannot \emph{produce} information.

We therefore have built certain notions, which are \emph{independent of any theory}. 
The cause of no-cloning and no-deleting follows from some principles, and whether or not
two copies and one copy have the same information content.

If the two quantities (information content of 
two copies and that of a single copy)
 differ in some theory, then we will have no-cloning and no-deleting 
in such a theory.

Now whether two copies and one copy actually \emph{have} the same information content, depends on 
specific theories.

In the classical case (i.e.$\sim$  when the underlying physical system 
consists of only distinguishable objects), two copies and one copy have the same information content. 
Suppose that a source will produce 
\(0\) if a certain team wins in a certain match. It produces \(1\) otherwise. 
%with probabilities \(p\) and \(1-p\) respectively, has the same information content
This source is \emph{equally informative} 
as a source producing \(00\) for ``win'', and \(11\) otherwise.
% probabilities \(p\) and \(1-p\). 
Here \(0\) and \(1\) are 
(classical) binary digits, encoded as two distinguishable objects (black and white balls, for example).

Below we see that in the quantum case, these quantities (information content of two copies and that 
of a single copy) are in general different, proving that one cannot clone or delete in this case. 
Specifically, in the quantum case, two copies will be seen to contain more information! 
%A good example is a photon 
%polarized either horizontally or at 60 degrees or minus 60 degrees from the horizontal, 
%each possibility being equally likely.Indeed for two identical copies of this state, there 
%is a measurement that will provide more than twice the expected amount of information than what 
%can be obtained with the optimal measurement on a single cope. 
Here by ``quantum case'', we only assume that there exist objects which are not perfectly distinguishable, and 
that they form a Hilbert space. We do not assume anything about the dynamics of this ``quantum case'',
except what is given by item (ii) above. Specifically, we assume that any dynamics \(\Lambda\) takes 
the stochastic mixture \(\sum_i p_i \varrho_i\) into \(\sum_i p_i \Lambda(\varrho_i)\). Note that 
there is nothing ``quantum'' about this notion of linearity. Such linearity is assumed, for 
example, in classical mechanics and classical electrodynamics.

To proceed, we must now obtain some numbers, the information contents of 
single copy and two copies, and compare them. So we must now define what we mean by information. 
It is well known that under certain natural axioms, one obtains entropy as a measure 
%of disorder and 
of information. 
%Note that subjectively, entropy measures disorder, while objectively, it measures information. 

%We now show that the no-cloning principle is implied by the above principle. 
%In this purpose suppose, that Alice prepares the state............

Let us recall briefly no-cloning and no-deleting statements.
%in an unknown  state 
%\(\left|\psi\right\rangle\), one cannot
%obtain
%\[
%\left|\psi\right\rangle \left|0\right\rangle \rightarrow 
%\left|\psi\right\rangle \left|\psi\right\rangle,
%\]
%even for open systems. Considering the environment inside the dynamics, this 
%states that one cannot obtain 
%\[
%\left|\psi\right\rangle \left|0\right\rangle \left|0\right\rangle_E \rightarrow 
%\left|\psi\right\rangle \left|\psi\right\rangle \left|e_{\psi}\right\rangle_E.
%\]
%
%Again a sharper formulation is possible, viz. 
Given a state from two (nonidentical) nonorthogonal states \(\left|\psi_1\right\rangle \) 
and \(\left|\psi_2\right\rangle \), \(\left\langle \psi_1 | \psi_2 \right\rangle \ne  0,1\)
%that 
one cannot have 
\begin{eqnarray}
\label{nonorthogonal_cloning}
&&\left|\psi_1\right\rangle \left|0\right\rangle \rightarrow 
\left|\psi_1\right\rangle \left|\psi_1\right\rangle, \nonumber \\
&&\left|\psi_2\right\rangle \left|0\right\rangle  \rightarrow 
\left|\psi_2\right\rangle \left|\psi_2\right\rangle ,
\end{eqnarray}
even for open systems. (A physical system is said to be open, if discarding part of the 
system is allowed along with (i), and (ii) stated above.) This is called no-cloning.
%for nonorthogonal \(\left|\psi_1\right\rangle \) and \(\left|\psi_2\right\rangle \).
Considering the environment inside the dynamics,
%Including the environment, 
this implies that 	
\begin{eqnarray}
\label{nonorthogonal_cloning_environ}
&&\left|\psi_1\right\rangle \left|0\right\rangle \left|0\right\rangle_E \rightarrow 
\left|\psi_1\right\rangle \left|\psi_1\right\rangle \left|e_{\psi_1}\right\rangle_E, \nonumber \\
&&\left|\psi_2\right\rangle \left|0\right\rangle \left|0\right\rangle_E \rightarrow 
\left|\psi_2\right\rangle \left|\psi_2\right\rangle \left|e_{\psi_2}\right\rangle_E,
\end{eqnarray}
is not possible by a single evolution.

Assuming a unitary dynamics, this statement can be proven. 
 This was done in 
Refs. \cite{18Zurek,19Dieks,20Yuen}, 
and was called the no-cloning \emph{theorem}. 
(Wigner was probably the first, who considered the
cloning problem within the quantum formalism \cite{ Wigner}.)

%Given two qubits (two-dimensional quantum systems, say a photon),
% from among two nonorthogonal states,
 %it is not possible to delete 
%a copy against another when given from among two nonorthogonal states. More formally,
Further,
the evolution
\begin{eqnarray}
\label{nonorthogonal_deleting}
&&\left|\psi_1\right\rangle \left|\psi_1\right\rangle \rightarrow 
\left|\psi_1\right\rangle \left|0\right\rangle, \nonumber \\
&&\left|\psi_2\right\rangle \left|\psi_2\right\rangle \rightarrow 
\left|\psi_2\right\rangle \left|0\right\rangle
\end{eqnarray}
is not possible for closed systems,
where 
\(\left\langle \psi_1 | \psi_2 \right\rangle \ne  0,1\).
(A physical system is called closed if we are not allowed to discard a part of the physical system.)
 This is called no-deleting.

Again the above statement of no-deleting  can be proven as a theorem, by assuming
a unitary evolution
%. This was done in 
%\cite{Braunstein_Pati_deleting}, 
and was 
called the no-deleting \emph{theorem}\cite {21Pati,22Pati}. Note that from a 
formal point of view, it is  not clear whether there is any
common physical origin of the above theorems or a relationship between them.

In the following, we raise both the no-cloning  and no-deleting  theorems to  principles. 
%We assume only linearity 
%in the dynamics. 
In particular, we do not assume that the dynamics is unitary.
 %and look for its connections with 
%thermodynamics.

We will now  show that in  physical systems (as specified before),
\begin{enumerate}
\item Conservation of   information (actually, no-\emph{increase} of  information) implies 
the no-cloning principle.
\item Conservation of  information (actually, no-\emph{decrease} of  information) implies 
the no-deleting principle.
\end{enumerate}
In the first implication, we allow discarding part of the system (tracing out) as a valid operation.

%\subsubsection{No-increase of entanglement implies the no-cloning principle}

%We now show that the no-cloning principle is implied by the principle that 
 %entanglement cannot increase 
%in bipartite systems under local operations (see Ref.\(^{4}\) in this regard).
% (see \cite{HH_basic} in this regard).

%\subsubsection{No-decrease of 
%entanglement in a closed system implies the no-deleting principle}

We will first show that the no-cloning principle is implied by the law of conservation of information. We 
will prove this by contradiction.
Suppose therefore that cloning is possible. That is, the transformation  in 
eq.$ \sim$ (\ref{nonorthogonal_cloning_environ}) is possible.
Then the average  input and output states are respectively
\begin{equation}
\varrho_{in} = \frac{1}{2}(\left|\psi_1\right\rangle \left|0\right\rangle \left|0\right\rangle 
\left\langle \psi_1 \right| \left\langle 0 \right| \left\langle 0 \right|
+ \left|\psi_2 \right\rangle \left|0\right\rangle \left|0\right\rangle 
\left\langle \psi_2 \right| \left\langle 0 \right| \left\langle 0 \right|),
\end{equation} and  
\begin{eqnarray}
\varrho_{out} &=& \frac{1}{2}(\left|\psi_1\right\rangle \left|\psi_1\right\rangle \left|e_{\psi_1}\right\rangle 
\left\langle \psi_1 \right| \left\langle \psi_1 \right| \left\langle e_{\psi_1} \right|  \nonumber \\
&+& \left|\psi_2 \right\rangle \left|\psi_2\right\rangle \left|e_{\psi_2}\right\rangle 
\left\langle \psi_2 \right| \left\langle \psi_2 \right| \left\langle e_{\psi_2}\right|),
\end{eqnarray}
where \(\left\langle \psi_1 |\psi_2 \right\rangle \neq 0,1\). Since 
\begin{equation} |\left\langle \psi_1 |\psi_2 \right\rangle | >
 |\left\langle \psi_1 |\psi_2 \right\rangle |^2 |\left\langle e_{\psi_1} |e_{\psi_2} \right\rangle|, 
\end{equation}
the von Neumann entropy of the average initial state  \(\varrho_{in}\) is less than 
%the entropy 
that of the average output state \(\varrho_{out}\), thus obtaining a violation of the law of conservation of 
information. 
(The von Neumann entropy of a state \(\varrho\), denoted as 
\(S(\varrho)\), is defined to be \(-\tr \varrho \log_2 \varrho\).)

We have therefore obtained that if cloning were possible, a single copy (on average) has less 
information than that in two copies. Note here the crucial importance of the quantum nature (nonorthogonality)
of the inputs. For classical inputs (orthogonal states), two copies have the same information as in a single
copy. The situation is similar for deleting, as we will see now.

We will now show that no-deleting principle is implied by the conservation of information, in particular, 
no-decrease of entropy.

Suppose that 
deletion is possible. Then one can effect the following evolution
(call it the ``deleting evolution'') in a closed system, for \(\left\langle \psi_1 | \psi_2 \right\rangle \ne  0,1\)
and a standard state \(\left|0\right\rangle\):
%The sharper form of the deleting, viz.
\begin{eqnarray}
&&\left|\psi_1\right\rangle \left|\psi_1\right\rangle \rightarrow 
\left|\psi_1\right\rangle \left|0\right\rangle,\nonumber \\
&&\left|\psi_2\right\rangle \left|\psi_2\right\rangle \rightarrow 
\left|\psi_2\right\rangle \left|0\right\rangle.
\end{eqnarray}
%for \(\left\langle \psi_1 | \psi_2 \right\rangle \ne  0\), also 
%We will show that violates the second law.
%This is because
Now \(\left|\psi_1\right\rangle \left|\psi_1\right\rangle \)
and \(\left|\psi_2\right\rangle \left|\psi_2\right\rangle \) are 
farther apart than \(\left|\psi_1\right\rangle \left|0\right\rangle \)
and \(\left|\psi_2\right\rangle \left|0\right\rangle\) and consequently 
the average input state 
\(\frac{1}{2}(\left|\psi_1\right\rangle \left|\psi_1\right\rangle 
\left\langle \psi_2 \right| \left\langle \psi_2 \right|
+ \left|\psi_2\right\rangle \left|\psi_2\right\rangle 
\left\langle \psi_1 \right| \left\langle \psi_1 \right|)\) 
has more von Neumann entropy than the average output state
\(\frac{1}{2}(\left|\psi_1\right\rangle \left|0\right\rangle 
\left\langle \psi_1 \right| \left\langle 0 \right|
+ \left|\psi_2 \right\rangle \left|0\right\rangle 
\left\langle \psi_2 \right| \left\langle 0 \right|)\). 
%Since the system is closed, 
%this decrease of entropy is therefore a violation of the conservative dynamics. 
So we obtain a violation of the law of conservation of information.
In other words, conservation of information implies the no-deletion principle.

One may also see deletion in the following way 
%\cite{unitary_deleting}
(which is again not possible under unitary dynamics 
\cite{22Pati}):
\begin{eqnarray}
&&\left|\psi_1\right\rangle \left|\psi_1\right\rangle \rightarrow 
                   \left|\psi_1\right\rangle \left|a_{\psi_1}\right\rangle, \nonumber \\
&&\left|\psi_2\right\rangle \left|\psi_2\right\rangle \rightarrow 
                     \left|\psi_2\right\rangle \left|a_{\psi_2}\right\rangle,
\end{eqnarray}
where \(\left\langle \psi_1 | \psi_2 \right\rangle \ne  0,1\), and 
\(\left|a_{\psi_1}\right\rangle\) and \(\left|a_{\psi_2}\right\rangle\)
are nearer than \(\left|\psi_1\right\rangle\) and \(\left|\psi_2\right\rangle\).
That is, \(|\left\langle \psi_1 | \psi_2 \right\rangle| < 
|\left\langle a_{\psi_1} | a_{\psi_2} \right\rangle|\). Again we see that 
the average input state 
\(\frac{1}{2}(\left|\psi_1\right\rangle \left|\psi_1\right\rangle 
\left\langle \psi_2 \right| \left\langle \psi_2 \right|
+ \left|\psi_2\right\rangle \left|\psi_2\right\rangle 
\left\langle \psi_1 \right| \left\langle \psi_1 \right|)\) 
has more von Neumann entropy than the average output state
\(\frac{1}{2}(\left|\psi_1\right\rangle \left|a_{\psi_1}\right\rangle 
\left\langle \psi_1 \right| \left\langle a_{\psi_1} \right|
+ \left|\psi_2 \right\rangle \left| a_{\psi_2}  \right\rangle 
\left\langle \psi_2 \right| \left\langle a_{\psi_2} \right|)\).

Therefore whatever is the form of the deleting principle, 
it violates the law of conservation of information.

%\section{Discussion}

Let us add here that it is conceivable that suitably 
extended forms of the  no-cloning and no-deleting  
principles would  imply  information conservation principle.

In conclusion, we have shown that  any theory for which dynamics is linear with respect to 
stochastic mixing,  
%\emph{independently of any theory},
%which incorporates the notion of measure of information \cite{24foot}, 
the  no-cloning  and no-deleting principles follow from the law of 
conservation of  information, and from whether two copies 
contain a different amount of information 
than a single copy. 
%This is independent of any theory. 
In particular, this result allows us to understand the physical
reason for which  perfect  cloning or deleting are impossible. They are forbidden
because they infringe a principle of conservation of information. 
Classically, two copies and one copy contain the same information. However in the quantum case, 
these information contents are generically different, putting restrictions on cloning and deleting processes.

In this context one can ask: What is the quantum formalism about? What laws are encoded in it?
Our result supports the view that the quantum formalism is just about information, which however 
is governed by some specific constraints,  making it more robust under processing.
In this spirit, one can consider the {\it laws of Nature as constraints} on processing of information.
These constraints have, in particular, their reflection in the properties of the classical \cite{Shannon}
and quantum information measures \cite{MPJ2003,MKPRJAU2003,Linden}.

\textbf{Acknowledgements}

The authors thank Robert Alicki for comments and 
Pawe{\l} and Karol Horodecki for discussions.
This work is supported by the BW grant of the 
University of Gda\'{n}sk and the Polish
Ministry of Scientific Research and Information Technology under the 
(solicited) grant No PBZ-Min-008/P03/2003,
 and EC grants RESQ Contract No. IST-2001-37559 
and QUPRODIS Contract No. IST-2001-38877. 
A.S.  and U.S. also acknowledge support from the Alexander von Humboldt Foundation.

\end{document}